\documentclass{aa} % for a referee version
\usepackage{graphicx}
\usepackage{natbib}
\usepackage{txfonts}
\bibpunct{(}{)}{;}{a}{}{,}
\begin{document}
\title{First signatures of strong differential rotation in A-type
  stars\thanks{Based on observations collected at the European
    Southern
    Observatory, La Silla}}

  \author{A. Reiners\inst{1}
    \and
    F. Royer\inst{2,3}}
  
  \offprints{A. Reiners}
  
  \institute{Hamburger Sternwarte, Universit\"at Hamburg,
    Gojenbergsweg 112, 21029 Hamburg, Germany\\
    \email{areiners@hs.uni-hamburg.de}
  \and
  Observatoire de Gen\`eve, 51 chemin des Maillettes, 1290 Sauverny, Switzerland \\ \email{frederic.royer@obs.unige.ch}
  \and
  GEPI, CNRS UMR 8111, Observatoire de Paris, 5 place Janssen, 92195 Meudon cedex, France
  }
  
  \date{Accepted 6 November 2003}
  \titlerunning{Differential rotation in A-stars}
  
  \abstract{We reanalyzed high quality spectra of 158 stars of
    spectral types A0--F1 and $v\,\sin{i}$ between 60 and
    150\,km\,s$^{-1}$. Using a Least Squares Deconvolution technique
    we extracted high $S/N$ broadening profiles and determined the
    loci of the Fourier transform zeros $q_{1}$ and $q_{2}$ where the
    $S/N$-ratio was high enough. The values of $v\,\sin{i}$ were
    redetermined and found to be consistent with the values derived
    by \cite{Royer02a}. For 78 stars $q_{2}$ could be determined and
    the ratio $q_{2}/q_{1}$ was used as a shape parameter sensitive
    for solar-like differential rotation (Equator faster than Pole).
    74 of the 78 stars have values of $q_{2}/q_{1}$ consistent with
    solid body rotation; in four of the 78 cases, values of
    $q_{2}/q_{1}$ are not consistent with rigid rotation. Although
    these stars may be binaries, none of the profiles shows signatures
    of a companion.  The Fourier transforms do not indicate any
    distortions and the broadening profiles can be considered due to
    single objects. One of those candidates may be an extremely rapid
    rotator seen pole-on, but for the other three stars of spectral
    types as early as A6, differential rotation seems to be the most
    plausible explanation for the peculiar profiles.
}
  
\maketitle

\section{Introduction}

The substantial difference between photospheres of solar-type stars
and A-type stars is the existence of a convective envelope. Due to the
ionization of hydrogen the cooler late-type stars harbour convective
envelopes where turbulent motions of the photospheric plasma can
occur. Stars of spectral types earlier than about F2 have no or only
very thin convective envelopes and properties of granular flows change
fundamentally.

The generally accepted activity paradigm places the stellar dynamo
believed to cause stellar activity at the boundary between the
convective envelope and the radiative core. Differential rotation
drives the dynamo action by winding up and amplifying the magnetic
flux tubes. The interaction of magnetic fields, differential rotation
and the convective envelopes are believed to be ultimately responsible
for stellar activity.

In a series of publications an onset of convection was searched for.
It is generally accepted that the onset of stellar activity occurs
between spectral types A7 and F5 depending on the observational
strategy. \cite{Wolff86} studied \ion{C}{ii} and \ion{He}{i} emission
and placed the onset of activity near $B-V = 0.28$, i.e. around
spectral type F0.  \cite{Schmitt97} concluded from X-ray data that
coronal emission is universal in the spectral range A7 to G9 implying
an onset of activity around spectral type A7.  Hotter stars are
expected to harbour shallow convective envelopes, these stars have
higher convective velocities which peak at about A3 until convection
disappears altogether at about A1 \citep{Renzini77}.  \cite{Gray89}
directly searched for the onset of convection analyzing line bisectors
of slowly rotating stars. In their targets the Doppler-shift
distribution of the granulation dominates the broadening of spectral
lines and a bisector reversal was found around spectral type F0.
Stronger asymmetries were found in the stars at the hot side of the
boundary indicating higher convective velocities.

Although stellar activity is not observed in early A-type stars, it is
not clear whether differential rotation may take place in early-type
stars. The absence of activity may simply reflect inefficient coupling
of surface magnetic fields and the lacking interface between the
radiative core and a convective envelope. There is no reason to
believe that rapidly rotating A-stars should rotate rigidly. In case
of the late-type Sun we know that the surface rotation law can be
approximated by
\begin{equation}
  \Omega(l) = \Omega_{\rm Equator} (1 - \alpha \sin^2{l}),
\label{DiffLaw}
\end{equation}
with $l$ the latitude and $\alpha_{\odot} \sim 0.2$ as derived from
Sun spots. In the solar case the Equator rotates about 20\% faster than
the Poles.  \cite{Gray77} searched for differential rotation in line
profiles of six A-stars finding no indications of differential
rotation within his error bars.

Also using line profiles, \cite{Reiners03} found signatures of
differential rotation in a sample of F-type stars. The earliest object
in their sample indicating differential rotation is of spectral type
F0IV/V. Applying the method used by \cite{Reiners03}, we search for
signatures of differential rotation in a large sample of A-star
spectra. The results are presented in the following.

\section{Observations and data analysis}

The spectra were observed with the ECHELEC spectrograph (ESO/La Silla)
and are part of a larger sample collected in the framework of an ESO
Key Programme. These observations were aimed at the determination of
fundamental parameters of early-type stars observed by HIPPARCOS
\citep{Gerbaldi89}. The total sample is described by \citet{Grenier99}
who measured radial velocities, and \citet[][ hereafter RGFG]{Royer02a}
who derived rotational velocities from these spectra.  The observed
spectral range spans from $4210$ to $4500$ \AA. The linear dispersion
is about $3.1$~\AA\,mm$^{-1}$, the slit width of 320~$\mu$m
corresponds to $1\farcs 52$ on the sky, and the resolving power is
about $28\,000$.

To search for the spectral signatures of stellar rotation laws,
broadening profiles were derived by applying a Least Squares
Deconvolution process (LSD). After constructing a $\delta$-template
comprising the strongest 150 lines taken from the Vienna Atomic Line
Database \citep{VALD} and according to stellar temperature, a
first-guess broadening profile was deconvolved using each pixel as a
free parameter in the fit. Since theoretical line depths match the
observational ones poorly, the equivalent widths of the incorporated
lines were optimized in a second step while leaving the broadening
profile fixed. During a few iterations the broadening profile and the
equivalent widths were optimized. Using this technique the spectral
lines are effectively deblended, the information contained in every
spectral line is used and the signal-to-noise ratio is significantly
enhanced. Consistency of the fit is checked by comparing theoretical
line depths to the derived ones.

Following \cite{Reiners02} we Fourier transformed the broadening
functions and measured the position of the first and second zeros
($q_{\rm 1}, q_{\rm 2}$). The ratio $q_{\rm 2}/q_{\rm 1}$ is a robust
observable for the shape of a rotational broadening function and a
direct indicator for solar-like differential rotation with the Equator
rotating faster than the Pole \citep[cp.][]{Reiners02}. We measured
the ratio $q_{\rm 2}/q_{\rm 1}$ for all stars the LSD procedure
yielded a stable and symmetric broadening function for.

The spectral quality used in this analysis in principle was sufficient
to follow the Fourier transformed broadening functions to the second
zero $q_{\rm 2}$ in stars with projected rotational velocities in the
range 60\,km\,s$^{-1} < v\,\sin{i} < 150$\,km\,s$^{-1}$.

\section{Results}

The outlined method was applied to the spectra of 158 stars of
spectral types A0--F1. The rotational velocity $v\sin{i}$ was derived
from the first zero $q_{\rm 1}$. For 78 of our sample stars the ratio
$q_{\rm 2}/q_{\rm 1}$ could be determined. For the disregarded 80
stars, either data quality was insufficient or the derived broadening
function showed obvious peculiarities probably due to binarity.

\begin{figure}
  \centering
  \resizebox{\hsize}{!}{\includegraphics{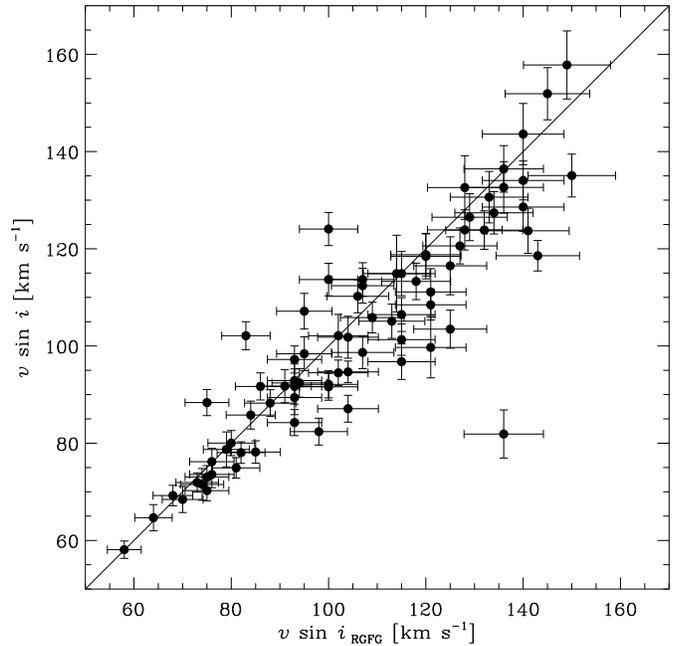}}
  \caption{\label{vsinicomp}Comparison of the $v\sin{i}$ values with
    the ones derived by RGFG.  The solid line is the one-to-one
    relation, and the linear regression between both data sets is
    given in Eq.~\ref{comp}.}
\end{figure}

The $v\sin{i}$ values are compared with the results from RGFG in
Fig.~\ref{vsinicomp}, for the subsample of 78 stars. RGFG also
determined their values of $v\,\sin{i}$ from $q_{1}$ but used selected
absorption lines while an ``overall'' broadening profile was
deconvolved here. The consistency of both scales is robustly estimated
using GaussFit \citep{Jes_98a,Jes_98b}. The resulting linear relation
is
\begin{equation}
\label{comp}
v\sin i = 0.99{\scriptstyle\pm 0.05}\,v\sin i_\mathrm{RGFG}-1.6{\scriptstyle\pm 4.6},
\end{equation}
both scales are in good agreement. The systematic differences occuring
due to the different ways obtaining the broadening profiles can be
neglected in our sample and should not be discussed here.

\begin{figure}
  \centering
  \resizebox{\hsize}{!}{\includegraphics[angle=-90]{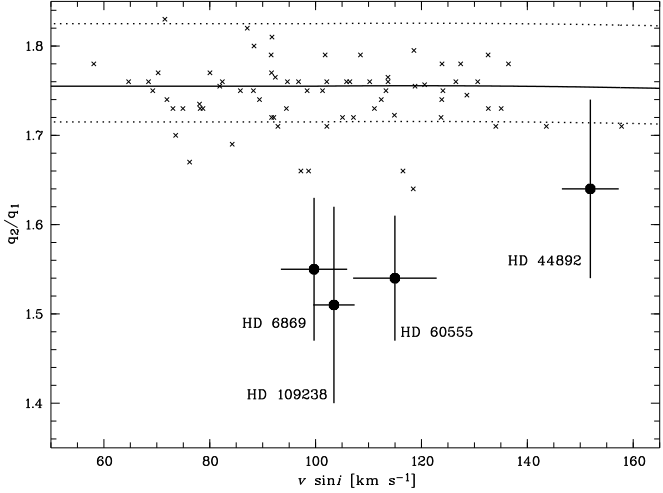}}
  \caption{\label{q2q1plot}Derived values of $q_{2}/q_{1}$ plotted against
    $v\, \sin{i}$ as derived from the first zero of the Fourier
    transform (with 1-$\sigma$ uncertainties). The region between
    dashed lines is consistent with solid body rotation for arbitrary
    limb darkening. For linear limb darkening with $\epsilon = 0.6$,
    $q_{2}/q_{1} = 1.76$ is expected (solid line). Crosses indicate
    results consistent with solid body rotation, typical errors are of
    the order of $\Delta q_{\rm 2}/q_{\rm 1} = 0.1$ (not plotted).
    Four stars not consistent with $q_{2}/q_{1} = 1.76$ are indicated
    by solid circles, error bars are plotted for them. }
\end{figure} 

We measured the second zeros of the Fourier transformed broadening
profiles to calculate the ratios $q_{2}/q_{1}$. The results are
plotted in Fig.\,\ref{q2q1plot}, typical errors are of the order of
$\Delta q_{\rm 2}/q_{\rm 1} \approx 0.1$.  A rigid rotator is expected
to yield a value of $q_{\rm 2}/q_{\rm 1}$ between $1.72$ and $1.83$
assuming a linear limb darkening law (indicated by dashed lines in
Fig.\,\ref{q2q1plot}). For the stars of our sample linear limb
darkening coefficients between 0.5 and 0.75 are expected during their
time on the Main Sequence \citep{Claret98}. Assuming a limb darkening
parameter of $\epsilon = 0.6$ rigid rotation would yield $q_{\rm
  2}/q_{\rm 1} = 1.76$ (solid line in Fig.\,\ref{q2q1plot}). The
results that are consistent with a value of $q_{\rm 2}/q_{\rm 1} =
1.76$ within the error bars are indicated by small crosses in
Fig.\,\ref{q2q1plot}, for the sake of readability no errors are
plotted for them. The second zero $q_{\rm 2}$ can only be determined
in spectra where the signal exceeds the noise level beyond $q_{\rm
  2}$, i.e. when a second sidelobe is detectable. In our case the
amplitude of the second sidelobe is at the noise level for many stars,
and these measurements of $q_{\rm 2}$ must be interpreted as lower
limits; thus some of the measurements of $q_{\rm 2}/q_{\rm 1}$ plotted
as crosses in Fig.\,\ref{q2q1plot} are essentially lower limits. For
74 of the 78 stars analyzed the broadening profiles are consistent
with solid body rotation.

%\begin{figure*}[!ht]
%  \resizebox{.3333\hsize}{!}{\includegraphics[angle=-90,clip=]{HD6869.ps}}
%  \resizebox{.3333\hsize}{!}{\includegraphics[angle=-90,clip=]{HD60555.ps}}
%  \resizebox{.3333\hsize}{!}{\includegraphics[angle=-90,clip=]{HD109238.ps}}
\begin{figure}[ht]
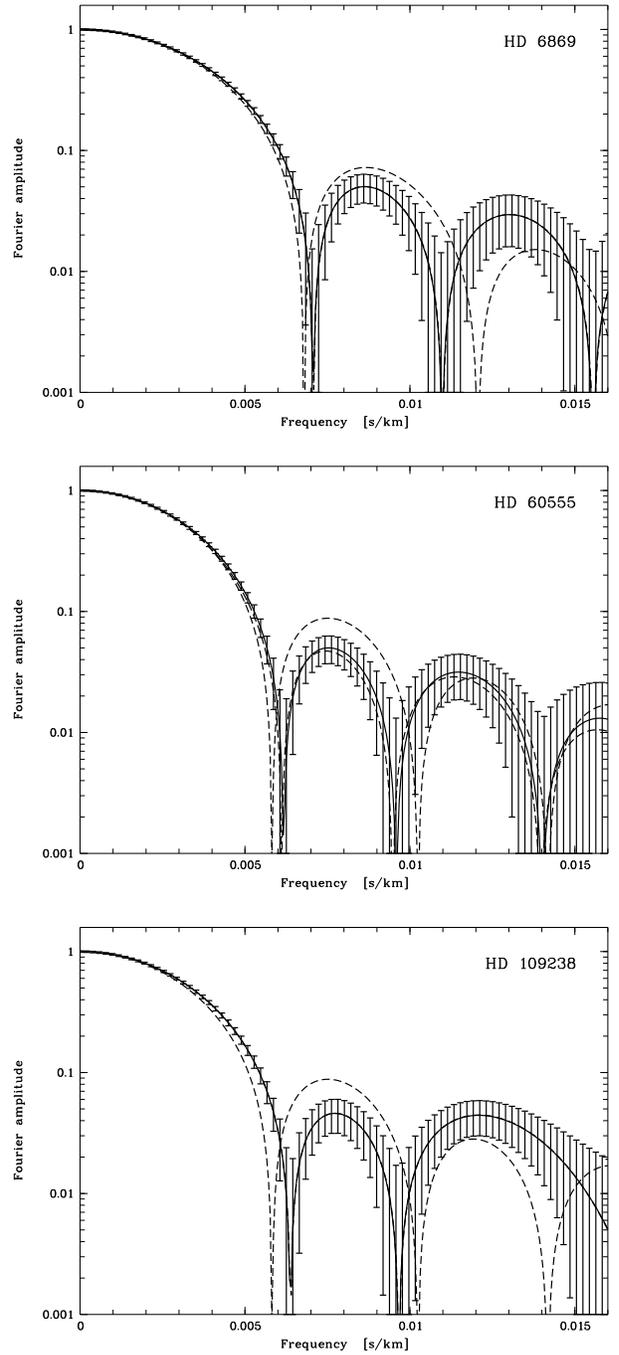

  \resizebox{\hsize}{!}{\includegraphics[angle=-90,clip=]{0175.f3a}}
  \resizebox{\hsize}{!}{\includegraphics[angle=-90,clip=]{0175.f3b}}
  \resizebox{\hsize}{!}{\includegraphics[angle=-90,clip=]{0175.f3c}}

  \caption{\label{FFTplots}Fourier transforms of HD~6869 (top panel), HD~60\,555 
    (middle), and HD~109\,238 (bottom panel) plotted with error bars.
    These stars show extremely small values of $q_{2}/q_{1}$. The
    spectra of HD~29\,920 and HD~145\,876 have values of $q_{2}/q_{1}$
    as expected for rigid rotators and are plotted with dashed lines
    for comparison in the top (HD~29\,902), middle and bottom (both
    HD~145\,876) panels, respectively.}
\end{figure}
%\end{figure*}

\begin{table*}[ht]
  \caption{\label{tab:Stars}Derived values of $v\,\sin{i}$ and $q_{2}/q_{2}$ for the four stars
    with values of $q_{2}/q_{1}$ significantly smaller than 1.76. Also
    given are strength of differential rotation in terms of $\alpha$ (cp.
    Eq.~\ref{DiffLaw}) and required values of equatorial velocities
    $v_{\rm e, rigid}$ and inclination angles $i$ if the value of
    $q_{2}/q_{1}$ is explained by rapid rotation seen pole-on (Sect.\,\ref{ultrafast}). Distances
    in pc, ROSAT X-ray luminosities and $v\,\sin{i}$ derived in RGFG are
    given in columns 8--10}
  \begin{tabular}{rlcccccrcc}
    \hline
    \hline
    \noalign{\smallskip}
    HD & Type & $v\,\sin{i}$& $q_{2}/q_{1}$ & $\alpha$ & $v_{\rm e, rigid}$ & $i$ & $d$ & $L_{\rm X}$ &$v\,\sin{i}$ \\
    &&[km\,s$^{-1}$]&&&[km\,s$^{-1}$]&& [pc] &[W m$^{-2}$]&(RGFG)\\  
    \noalign{\smallskip}
    \hline
    \noalign{\smallskip}
        6869 & A9V    & $100 \pm 6$ & $1.55 \pm 0.08$ & $0.28 \pm 0.10$ & (460) & (13\degr) &  87 & 570 & 121\\
     60\,555 & A6V    & $115 \pm 7$ & $1.54 \pm 0.07$ & $0.29 \pm 0.08$ & (470) & (14\degr) & 134 &     & 114\\
    109\,238 & F0IV/V & $103 \pm 4$ & $1.51 \pm 0.11$ & $0.32 \pm 0.13$ & (500) & (13\degr) & 133 &     & 125\\
    \noalign{\smallskip}
    44\,892 & A9/F0IV & $152 \pm 5$ & $1.64 \pm 0.10$ & $0.16 \pm 0.16$ & 400 &  22\degr  & 160 &     & 145\\
    \hline
  \end{tabular}
\end{table*}

Four of our measurements are not consistent with $q_{\rm 2}/q_{\rm 1}
= 1.76$, they are indicated by full circles in Fig.\,\ref{q2q1plot}
and errors bars are plotted for them. For three of those --- HD~6869
(A9V), HD~60\,555 (A6V) and HD~109\,238 (F0IV/V) --- the values of
$q_{\rm 2}/q_{\rm 1}$ are significantly smaller than 1.7. The fourth
star (HD~44\,892, A9/F0IV) has a value of $q_{\rm 2}/q_{\rm 1}$
marginally consistent with $q_{\rm 2}/q_{\rm 1} > 1.7$ within its
error bars. We will discuss this star in Sect.\,\ref{ultrafast}.

The Fourier transforms of HD~6869, HD~60\,555 and HD~109\,238 are
plotted with error bars in Fig.\,\ref{FFTplots}. Overplotted are the
Fourier transformed line profiles of stars with similar values of
$v\,\sin{i}$ that are consistent with rigid rotation ($q_{\rm
  2}/q_{\rm 1} = 1.76$). While different velocity fields, e.g.,
turbulence, may influence the amplitudes of the sidelobes, the zeros
of the Fourier transform arise from rotational broadening
\citep[cp.][]{Gray76}. One mechanism known to change the ratio $q_{\rm
  2}/q_{\rm 1}$ in the manner found in HD~6869, HD~60\,555 and
HD~109\,238 is solar-like differential rotation. The strength of
differential rotation in terms of the parameter $\alpha$ in
Eq.\,\ref{DiffLaw} can be calculated from $q_{\rm 2}/q_{\rm 1}$
\citep{Reiners03}, and the respective values of $\alpha$ are given in
Table\,\ref{tab:Stars} together with the spectral types and
$v\,\sin{i}$ of the four suspected differential rotators.

\subsection{Binarity}

The spectra we used were also studied in the aim of deriving radial
velocities by \cite{Grenier99}. The authors gave a ``shape'' flag for
the correlation functions for our four candidates:\\

\begin{tabular}[h]{ll}
-- HD~6869: & ``probable double'',\\
-- HD~44\,892: & ``probable double'',\\
-- HD~60\,555: & ``suspected double'',\\
-- HD~109\,238: & ``probable double''.\\
\end{tabular}\\

\noindent Using our deconvolution method, however, we find indication for
double peaks in the broadening functions of 23 of our targets, but not
in the spectra of these four stars. Note that this does not mean that
the 135 others are single stars since the luminous A-type stars
dominate spectra of, e.g., binaries consisting of an A-type and a
G-type star. The G-type spectrum will easily be hidden in the light of
the A-type star. To be complete we checked the shape of the
correlation function by cross-correlating our template with the
spectra and found no indications for binarity either.

For HD~44\,892 and HD~109\,238, the literature gives hints about a
single star status. There is no evidence of binarity for
\object{HD~44\,892}, neither in HIPPARCOS data nor in Speckle
observations \citep{Mason01}. It can be considered as single with a
high level of confidence, and its spectrum is surely not affected by
any significant contamination. \object{HD~109\,238} is part of the
sample observed by \citet{Abt95}. The MK classification they derive
for this object is F0V, with no suspicion of spectroscopic binarity.
 
For the two other objects, \object{HD~6869} and \object{HD~60\,555}
two spectra are available in our data set.  Individual observations of
HD~6869 and HD~60\,555 are separated by $383$\,d and $767$\,d,
respectively.  Inspection of the broadening functions derived from the
individual spectra yields no indication of variability due to relative
motions of binary components. Coadded spectra were used to derive the
values of $q_{\rm 2}/q_{\rm 1}$ for both stars.  Both targets are
indicated as binary stars in the literature and are discussed below.

\subsubsection{HD~6869} 
HD~6869 is a binary system seen with an angular separation of
about $1\farcs 2$ \citep{Hip,Horch96}. With such a low separation,
compared to the width of the slit, the collected spectra are expected
to be contaminated by the light of the secondary.  The individual
magnitudes in HIPPARCOS and TYCHO bands are given by
\citet{Fabricius00} for both components and are respectively:\\

\begin{tabular}[h]{cll}
-- & for the primary: & $Hp=7.190\pm 0.003$~mag,\\
   & & $B_T=7.42\pm 0.01$~mag,\\
\smallskip
   & & $V_T=7.16\pm 0.01$~mag,\\
-- & for the secondary: & $Hp=8.790\pm 0.014$~mag,\\
   & & $B_T=9.34\pm 0.01$~mag,\\
   & & $V_T=8.76\pm 0.01$~mag.\\
\end{tabular}\\

\noindent No significant interstellar absorption is expected \citep{Lucke78}
from the position of the star in the solar neighborhood ($l=294\degr$,
$b=-70\degr$, $d=87$\,pc). Considering no reddening, the temperatures
of each component, derived from the $B_T-V_T$ color index
(Appendix~~\ref{calib}), are respectively $7500$~K and $6000$~K,
according to Eq.~\ref{Teff}. The magnitude difference in $B_T$
corresponds to a flux ratio of about $5.5$ around $4450$~\AA, taking
into account the effective temperature of both components. Given the
angular separation and the large magnitude difference, the spectrum of
HD~6869 can be considered as a single star spectrum of the dominating
A-type star.

\subsubsection{HD~60\,555} 
This star has been very little studied. Its spectral type in the
catalogue published by \citet{Houk82} indicates a composite spectrum:
A5/7V+(F).  \citet{Grenier99} flagged this star as variable in radial
velocities from these two spectra. The ratio $E/I$ of external to
internal error on radial velocity is $2.93$.  The absorption spectrum
of HD~60\,555 in the observations does not show any evidence of
multiplicity, especially in the Fourier transform no indication of
contamination due to the light of a secondary is apparent.

\subsection{Extremely fast rotation}
\label{ultrafast}

Alternative to differential rotation the shape of the broadening
function and the value of $q_{\rm 2}/q_{\rm 1}$ can also be affected
by very rapid rotation and gravity darkening possibly observed pole-on
\citep[cp.][]{Reiners03a}.  Flux is redistributed from the line's
wings to the center when the Equator becomes cooler due to gravity
darkening. As far as the lines considered are not dominated by
temperature and gravity variations over the stellar surface --- which
is the case, e.g, in the weak lines of early A-type stars as shown by
\cite{Gulliver94} --- the value of $q_{\rm 2}/q_{\rm 1}$ is diminished
by this effect. According to \cite{Reiners03a}, the ratio $q_{\rm
  2}/q_{\rm 1}$ then only depends on the equatorial velocity $v_{\rm
  e}$ and on the gravity darkening law. We assume a linear gravity
darkening law according to \cite{Claret98} and calculate the
equatorial velocities $v_{\rm e, rigid}$ required to produce the
measured values of $q_{\rm 2}/q_{\rm 1}$ assuming solid body rotation
for the four suspected differential rotators. The results and the
respective inclination angles $i$ are given in columns six and seven
of Table\,\ref{tab:Stars}.

For HD~6869, HD~60\,555 and HD~109\,238 the values of $v_{\rm e,
  rigid}$ are larger than breakup velocity; for these stars rapid
solid body rotation can be ruled out as the mechanism solely
responsible for the diminished ratio $q_{\rm 2}/q_{\rm 1}$. In case of
HD~44\,892 the rotational velocity required for the measured ratio
$q_{\rm 2}/q_{\rm 1} = 1.64$ is of the order of breakup velocity. Thus
differential rotation as well as rapid solid body rotation are the two
possible explanations for the measured profile shape of HD~44\,892.

\section{Conclusions}

We reanalyzed high quality data formerly discussed by \cite{Grenier99} 
and \cite{Royer02a}, with the aim of searching for differential
rotation in early-type stars. With an iterative Least Squares
Deconvolution method we obtained high quality broadening profiles of
158 stars with projected rotational velocities in the range
60\,km\,s$^{-1} < v\,\sin{i} < 150$\,km\,s$^{-1}$. We disregarded the
profiles of 80 of them due to obvious asymmetries or multiplicity. For
78 stars the broadening profiles apparently reflect the rotational
broadening law. Profile distortions were analyzed in terms of the
ratio of the first two zeros of the Fourier transform $q_{\rm
  2}/q_{\rm 1}$. Within the errors, 74 of the 78 measured profiles are
consistent with the assumption of rigid rotation. Due to data quality
many measurements must be considered lower limits and from this sample
no conclusion can be drawn concerning values of $q_{\rm 2}/q_{\rm 1}$
possibly larger than 1.8.

Unfortunately, for none of our sample stars interferometric
measurements are avaliable to our knowledge, e.g., Altair and Sirius
have rotational velocities outside our range determined by data
quality, and they are not contained in our sample.

Four stars are analyzed in detail, the profile of the A9/F0IV star
HD~44\,892 is only marginally consistent with rigid rotation. It is
likely that its profile is distorted either by differential rotation
or by very rapid rotation seen pole-on; in the latter case HD~44\,892
would be the first star that directly shows signatures of gravity
darkening in mean profile broadening as proposed by \cite{Reiners03a}.
Comparison to interferometric results would be interesting especially
for this star.

The broadening functions of the three stars HD~6869 (A9V), HD~60\,555
(A6V) and HD~109\,238 (F0IV/V) are not consistent with rigid --- even
very rapid --- rotation since their equatorial velocities would be
larger than breakup velocity. Although some authors suspect these
stars being binaries, in our high quality spectra we find no
indications of multiplicity neither in data nor in Fourier space.
Since contamination due to secondaries are easily visible in Fourier
space --- where no sharp zeros should occur in case of the profile
being a sum of two --- we consider the spectra single star spectra.
Differential rotation seems to be the most plausible explanation for
the observed profile distortions. For these three stars the Equator is
rotating about 30\% faster than the polar regions. Thus we conclude
that significant differential rotation seems to take place even in
early-type stars not harbouring deep convection zones, the earliest
object is the A6 dwarf HD~60\,555.

If differential rotation is the driving mechanism for stellar
activity, these stars should be active, too. X-ray emission from
HD~6869 was detected with the ROSAT mission and the other stars may
also be X-ray sources but were simply too far away for a detection.
Whether differential rotation is a common phenomenon in these stars
cannot be answered by this work since only very strong differential
rotation is detectable with our method. The finding of strong
differential rotation among A-type stars indicates that there is no
abrupt change in rotational laws of stars around the boundary where
surface convection sets in.

\begin{acknowledgements}
  A.R. acknowledges financial support from Deutsche
  Forschungsgemeinschaft DFG-SCHM 1032/10-1. This work was partly
  supported by the Swiss National Science Foundation.
\end{acknowledgements}

\appendix
\section{\boldmath{$B_T-V_T$} versus \boldmath{$T_\mathrm{eff}$}}
\label{calib}
The effective temperature can be approximated using the color index
$B_T-V_T$. This calibration is built from the catalogue given by
\citet{Cayrel97}. They compile [Fe/H] and $T_\mathrm{eff}$
determinations from the literature. These data together with TYCHO
colors for these stars allow the calibration of $T_\mathrm{eff}$ as a
function of $B_T-V_T$. The resulting formula is:
 
\begin{eqnarray}
\label{Teff}
\log(T_\mathrm{eff}) &=& 4.0132 - 0.87573\,x + 2.2194\, x^2 \\ \nonumber 
                     & & - 5.0087\, x^3 + 6.7676\, x^4 - 5.1069 \,x^5  \\ \nonumber
                     & & + 2.0638\, x^6 - 0.39629\, x^7 + 0.024548\, x^8,
\end{eqnarray} with $x = B_T-V_T$.

\end{document}